# Electro-Magnetic Decoupling Preconditioner for Eddy Current Problems with External Circuit Coupling

S. Hiruma[1], T. Mifune[2], and T. Matsuo[2]

[1]Graduate School of Information Science and Technology, Hokkaido University, Hokkaido, 060-0814, Japan
[2]Graduate School of Engineering, Kyoto University, Kyoto, 615-8510, Japan

**This paper proposes an efficient and scalable preconditioning strategy for eddy current problems involving coupled external circuits. The approach, named Electro-Magnetic Decoupling (EMD) preconditioner, decomposes the discrete system into vector and scalar potential components and applies tailored preconditioners to each. In particular, strong preconditioning is applied to the scalar component to address the spectral degradation induced by the discrete Laplacian. The method was evaluated on four eddy current models with varying frequencies, conductor topologies, and excitation types. Compared to the conventional incomplete Cholesky preconditioner, the EMD approach achieved up to 20 times fewer iteration counts and up to 20 times faster iterative solver time. Moreover, the method supports physics-level parallelization, allowing efficient treatment of independently excited conductor domains. The EMD framework is compatible with algebraic multigrid, domain decomposition, and direct solvers, offering flexibility and robustness for large-scale electromagnetic simulations.**

## I. Introduction

The finite element method (FEM), particularly when using edge elements, has become an indispensable tool for analyzing eddy current phenomena in electrical machinery and power devices. Accurate prediction of eddy current distributions is critical in the design and optimization of such systems, as eddy current losses can significantly impact efficiency, thermal performance, and reliability. For instance, in passive components driven by high-frequency power sources, such as those employing pulse width modulation (PWM), eddy current losses in the windings and surrounding conductive structures can lead to substantial energy dissipation. Similarly, in high-speed rotating electrical machines, alternating magnetic fields induce eddy currents in the conductors and structural components, resulting in alternating current (AC) losses. These losses include those caused by the skin and proximity effects in windings, as well as eddy current losses in permanent magnets. Capturing these loss mechanisms with high fidelity requires advanced numerical modeling techniques that can accurately resolve both geometric complexities and material inhomogeneities.

In current practice, the finite element analysis of eddy current problems widely relies on iterative solvers based on the conjugate gradient (CG) method combined with incomplete Cholesky (IC) decomposition as a preconditioner. The ICCG approach is highly robust and versatile, offering reliable convergence across a broad range of problems. Furthermore, the use of parallelization techniques such as localized incomplete Cholesky (localized IC) [1] enables theoretically high parallel efficiency, making ICCG particularly well-suited for modern large-scale computational architectures.

In recent years, mathematically sophisticated preconditioning techniques have attracted attention, including algebraic multigrid (AMG) methods [2],[3], auxiliary space preconditioners (ASP) [4], and domain decomposition methods (DDM) [5]. Each of these approaches provides notable advantages: AMG is known for its ability to accelerate convergence in curl-curl type problems, ASP leverages auxiliary spaces to efficiently handle H(curl) or H(div) formulations, and DDM provides a natural framework for parallelization by dividing the domain into smaller subproblems. However, practical limitations remain. AMG often faces challenges in achieving high parallel efficiency on unstructured meshes, which are common in complex eddy current simulations. ASP is primarily formulated for time-domain problems, and its effectiveness for frequency-domain analyses has not been extensively studied. Although DDM holds promise, applying it effectively to edge-element-based finite element formulations requires special considerations [6], and published reports demonstrating its practical success for such cases are still limited.

Consequently, despite the existence of more advanced methods, IC-based preconditioning continues to be widely used in industrial eddy current simulations, owing to its simplicity and robustness. Yet, when external circuits are coupled to the eddy current model and eddy currents are induced in windings, the performance of IC preconditioners deteriorates significantly, often resulting in a dramatic increase in iteration count. Addressing this issue is therefore of great importance for enabling efficient and scalable eddy current simulations under such challenging conditions.

Motivated by these challenges, this study aims to develop a novel and effective preconditioning strategy for eddy current problems with coupled external circuits. We propose the electro-magnetic decoupling (EMD) preconditioner, which is designed to separately address the magnetic and electric components of the system matrix. The core idea of the EMD preconditioner is to apply distinct preconditioning strategies to the magnetic vector potential and the electric scalar potential components, thereby improving the condition number of the system while minimizing additional computational overhead. By tailoring the strength of the preconditioners for each component, the method effectively mitigates the convergence issues caused by poor conditioning, particularly those associated with the coupling of external circuits and elongated conductor geometries.

The contributions of this work are threefold. First, we identify the root cause of the poor conditioning observed in eddy current problems coupled with external circuits. Second, we propose a block-structured preconditioning strategy that accelerates iterative solvers by separately treating the vector and scalar potential components. Third, we provide comprehensive numerical experiments demonstrating that the proposed method significantly reduces the iteration count and computational time compared to conventional IC-based solvers, across a range of realistic test cases including models with complex winding geometries and skin-effect-resolving meshes.

The remainder of this paper is organized as follows. Section II formulates the eddy current problem and its finite element discretization. Also, it details the design and theoretical basis of the proposed preconditioner. Section III presents the numerical experiments and results, and Section IV concludes the paper with remarks on future work.

## II. Mathematical Formulation

### A. Governing Equations

We consider the time-harmonic eddy current problem formulated using the magnetic vector potential $\boldsymbol{A}$ and the electric scalar potential $\varphi$ ($A$-$\varphi$ formulation). Let $\Omega \subset \mathbb{R}^3$ be a bounded computational domain comprising conducting $\Omega_c$ and non-conducting $\Omega \setminus \Omega_c$ subdomains. The governing Maxwell's equations reduce to:

$$\nabla \times (\nu \nabla \times \boldsymbol{A}) + j\omega\sigma\boldsymbol{A} + \sigma\nabla\varphi = 0, \qquad \text{in } \Omega \tag{1}$$

$$\nabla \cdot \sigma(j\omega\boldsymbol{A} + \nabla\varphi) = 0, \qquad \text{in } \Omega_c \tag{2}$$

where $\nu, \sigma, \omega$ and $j$ denote the reluctivity, conductivity, angular frequency and imaginary unit, respectively. Eq. (1) represents Ampère's law with Ohm's law substitution, and (2) enforces current continuity.

### B. Excitation Through Boundary Condition

In this paper, we focus on voltage and current excitation through the exterior boundary conditions and local excitation, which is also referred to as contacts at $\partial\Omega$ [7].

#### 1) Voltage Excitation

Voltage excitation is implemented by imposing Dirichlet conditions on the scalar potential $\varphi$

$$\varphi = U_k, \qquad \text{on } \Gamma_k \tag{3}$$

where $U_k$ is the voltage applied to the k-th port surface $\Gamma_k$.

The weak formulation of (1) and (2) considering the voltage excitation (3) yields:

$$\int_\Omega \nu\nabla \times \boldsymbol{A} \cdot \nabla \times \boldsymbol{A}' d\Omega + j\omega \int_{\Omega_c} \sigma\boldsymbol{A} \cdot \boldsymbol{A}' d\Omega + \int_{\Omega_c} \sigma\nabla\varphi \cdot \boldsymbol{A}' d\Omega = -\sum_k \left( \int_{\Omega_c} \sigma\nabla\tilde{v}_k \cdot \boldsymbol{A}' d\Omega \right) U_k \tag{4}$$

$$\int_{\Omega_c} \sigma\boldsymbol{A} \cdot \nabla\varphi' d\Omega + \frac{1}{j\omega}\int_{\Omega_c} \sigma\nabla\varphi \cdot \nabla\varphi' d\Omega = -\sum_k \left( \frac{1}{j\omega}\int_{\Omega_c} \sigma\nabla\tilde{v}_k \cdot \nabla\varphi' d\Omega \right) U_k \tag{5}$$

where $\boldsymbol{A}', \varphi'$ are the test functions for magnetic vector potential and electric scalar potential, and $\tilde{v}_k$ is an arbitrary scalar function that is 1 on $\Gamma_k$ and 0 on $\partial\Omega \setminus \Gamma_k$.

#### 2) Current Excitation

Current excitation is imposed via integral constraints on the normal component of the conduction current

$$\int_{\Gamma_k} \sigma(j\omega\boldsymbol{A} + \nabla\varphi) \cdot \boldsymbol{n} dS = -I_k, \tag{6}$$

where $I_k$ is the net current injected into the domain through port $\Gamma_k$. The weak formulation (4) and (5) are simultaneously solved with the following equation derived from (6):

$$\int_{\Omega_c} \sigma(j\omega\boldsymbol{A} + \nabla\varphi) \cdot \nabla\tilde{v}_k d\Omega + U_k \int_{\Omega_c} \sigma|\nabla\tilde{v}_k|^2 d\Omega = -I_k \tag{7}$$

### C. Finite Element Discretization and Circuit Coupling

The system is discretized using edge (Nédélec) elements for $\boldsymbol{A}$, and Lagrange nodal elements for $\varphi$. This results in a mixed finite element formulation coupling edge-based and node-based degrees of freedom.

The finite element equation considering the voltage excitation yields:

$$\begin{bmatrix} C^\top M_\nu C + j\omega M_\sigma & M_\sigma G \\ G^\top M_\sigma & \frac{1}{j\omega} G^\top M_\sigma G \end{bmatrix} \begin{bmatrix} a \\ \phi \end{bmatrix} = \sum_k \begin{bmatrix} b_{k1} \\ b_{k2} \end{bmatrix} U_k \tag{8}$$

where $C, G, M_\nu, M_\sigma, a, \phi, b_{k1}$ and $b_{k2}$ are the discrete curl operator matrix, discrete gradient operator matrix, reluctivity matrix, conductivity matrix, vector of unknown coefficients for $\boldsymbol{A}$, vector of unknown coefficients for $\varphi$ and contributions of impressed sources and boundary conditions, respectively.

In the case of current excitation, (8) is extended by incorporating the constraint equation (7), which links the net current to the potentials via integral conditions.

The coupling with external circuits is realized by directly linking the excitation variables $U_k$ and $I_k$ to circuit equations. Specifically, $U_k$ and $I_k$ are treated as interface variables that interact with the external circuit through algebraic coupling, enabling bidirectional interaction between the finite element model and external lumped-element circuits.

### D. Preconditioner Design

#### 1) Motivation and Theoretical Basis

In eddy current problems with external circuit coupling, the system matrix arising from the finite element discretization often exhibits a large condition number compared with that in magneto-static problems, which severely deteriorates the convergence rate of iterative solvers. This is primarily caused by the presence of small singular values in the matrix $G^\top M_\sigma G$, the discrete Laplacian associated with the continuity equation in the conducting domain. Such small singular values are particularly pronounced in cases involving long conductors or specific mesh configurations. They significantly increase the condition number and thus degrade the performance of conventional preconditioned iterative solvers.

Although the singular values of the discrete Laplacian can, in principle, be related to the analytical eigenvalues of Laplace problems [8], these results are well known and are therefore omitted here for brevity. The key point is that the poor conditioning originates from the electric scalar potential component, which necessitates a preconditioning strategy that targets this specific part of the system.

To address this, we propose applying separate preconditioners to the magnetic vector potential and the electric scalar potential components. This approach efficiently mitigates the convergence issues caused by poor conditioning while maintaining computational efficiency.

#### 2) Structure of the EMD preconditioner

The proposed electro-magnetic decoupling (EMD) preconditioner is designed to exploit the block structure of the system matrix. It is defined as:

$$M = \begin{bmatrix} M_1 & O \\ O & M_2 \end{bmatrix} \tag{9}$$

where $M_1$ is the preconditioner applied to the magnetic vector potential component associated with the matrix $C^\top M_\nu C + j\omega M_\sigma$, and $M_2$ is the preconditioner applied to the electric scalar potential and circuit unknowns associated with the matrix $\frac{1}{j\omega} G^\top M_\sigma G$ and constraint and circuit equations.

The design guidelines for the EMD preconditioner are as follows:

- The first preconditioner $M_1$ can be computationally inexpensive, such as the IC preconditioner. This is important because the matrix corresponding to the magnetic vector potential component is generally large, and minimizing its computational cost is essential.
- The second preconditioner $M_2$ can be stronger, chosen to efficiently increase the lower bound of the nonzero minimum singular value of the system. Suitable choices for $M_2$ include AMG methods, DDM or direct solvers. Since the number of degrees of freedom for the scalar potential (i.e., the number of nodes) is typically much smaller than that for the vector potential (i.e., the number of edges), the overall increase in computational cost remains limited even when employing a stronger preconditioner for $M_2$.

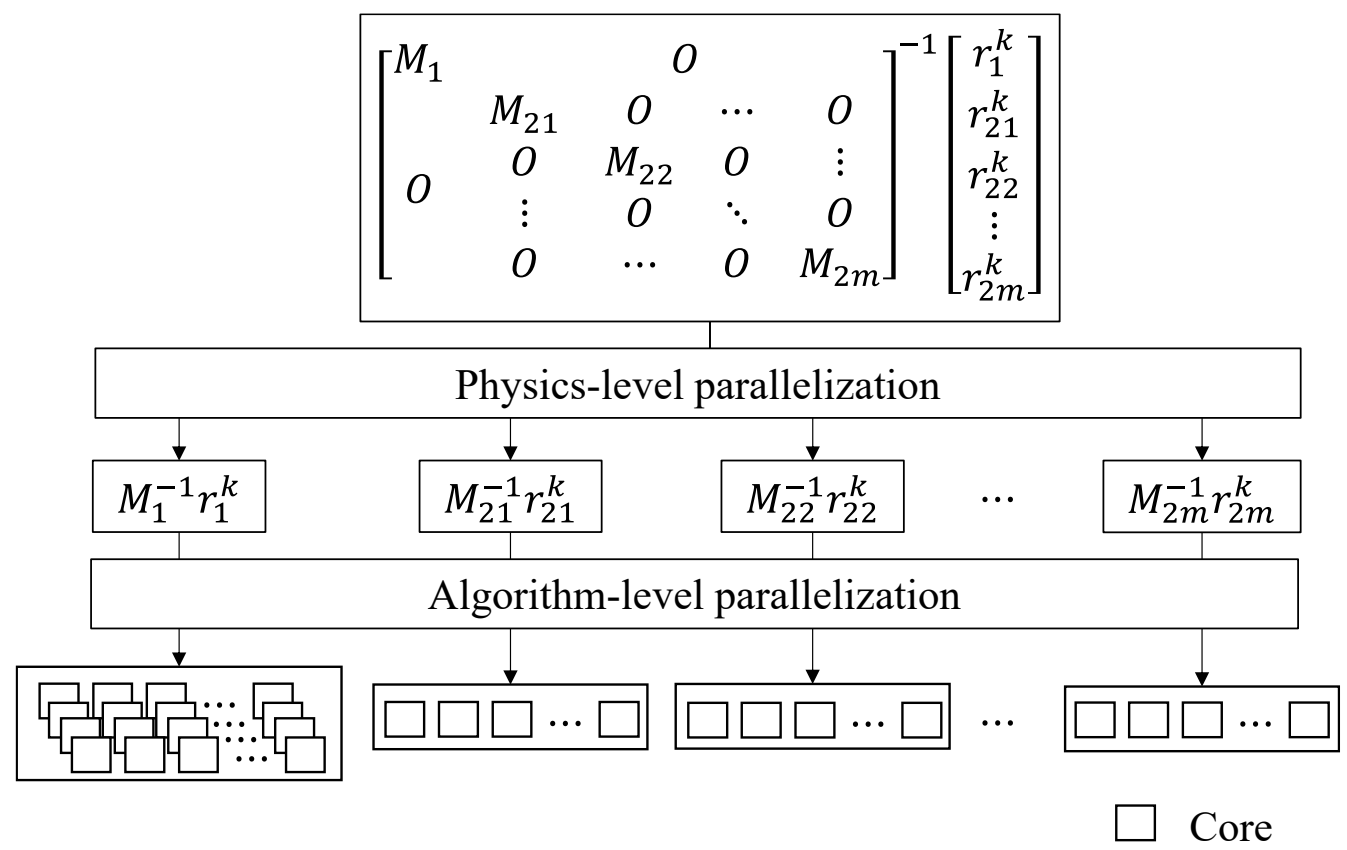

Fig. 1. Schematic illustration of the parallelization structure of the proposed EMD preconditioner. The upper layer shows physics-level parallelization, where the magnetic vector potential ($M_1$) and electric scalar potential ($M_2$) components are processed independently, and $M_2$ is further parallelized across independent conductor regions. The lower layer shows algorithm-level parallelization, where each preconditioner ($M_1$, $M_2$ blocks) is parallelized through domain decomposition or other numerical techniques. This dual-level design enables efficient use of modern high-performance computing architectures.

By selecting appropriate preconditioners for $M_1$ and $M_2$, the EMD approach can achieve substantial improvement in convergence while keeping additional computational overhead to a minimum.

#### 3) Implementation Considerations

The block-diagonal structure of the EMD preconditioner allows the preconditioners for the magnetic vector potential ($M_1$) and the electric scalar potential ($M_2$) to be applied independently and in parallel. This separation is based on the distinct physical roles of these components and can thus be regarded as a form of physics-level parallelization. This design facilitates high parallel efficiency on modern computational architectures, including non-uniform memory access (NUMA)-based systems and multi-node clusters, by enabling flexible load balancing and scalability.

Moreover, the structure of $M_2$ offers an additional physics-level parallelization. In practical electromagnetic devices, coil regions typically consist of multiple independent conductor regions. Since the continuity equation applies independently within each conductor, the discrete Laplacian $G^\top M_\sigma G$ associated with the scalar potential exhibits a block-diagonal structure across these conductor regions. Consequently, $M_2$ inherits this block-diagonal structure such that $M_2 = \text{blockdiag}(M_{21}, M_{22}, \dots, M_{2m})$, where $m$ is the number of the independent conductor regions, and can additionally be parallelized across conductor regions.

This dual-level physics-based parallelization — first between $M_1$ and $M_2$, and second within $M_2$ across conductor regions — complements algorithm-level parallelization methods such as domain decomposition or computational domain partitioning, which are widely used in localized IC and other preconditioners. These combined approaches facilitate the proposed preconditioner to fully exploit the parallel processing capabilities of modern hardware as shown in Fig. 1.

### E. Relation to Existing Methods and Novelty

The proposed EMD preconditioner can also be interpreted within the framework of the $A$ formulation as a two-level additive Schwarz preconditioner, expressed as:

$$M^{-1} = M_1^{-1} + (j\omega)^{-2} G M_2^{-1} G^{\top}. \tag{10}$$

This form of preconditioner is not entirely new; Hiptmair's hybrid smoother for edge-element-based multigrid methods [9] can be written in a same structure. In the context of Hiptmair's original work, both preconditioners corresponded to smoothers based on Gauss-Seidel relaxation, reflecting the focus on general eddy current problems and multigrid convergence acceleration.

The novelty of the present work lies in the specific design principle applied to the EMD preconditioner. Unlike previous studies where both levels of the preconditioner employed similar smoothing strategies, we propose a targeted approach. This design is motivated by the specific numerical challenges arising in eddy current problems with external circuit coupling, where the continuity equation component is the primary source of poor conditioning. By explicitly addressing this source with a stronger preconditioner for $M_2$, the proposed method provides a practical and effective solution distinct from conventional hybrid smoothers or multigrid-based preconditioners. This tailored design enables substantial improvements in convergence without introducing prohibitive computational cost.

## III. Numerical Result

### A. Model Description

#### 1) Model A: 21-turn coil model

This model consists of 21 turns wound around a magnetic core as shown in Fig. 2. It is excited by a 1 V voltage source at 50 Hz with one input and one output port defined. The winding material is copper with a conductivity of $5.96 \times 10^7$ S/m, and the magnetic core is assumed to be linear with a relative permeability of 1000. The mesh contains 1,441,102 edge DOFs in $\Omega$ and 130,811 conducting node DOFs in $\Omega_c$.

#### 2) Model $A_p$: Parametric elongated coil model

Coils with $N_{\text{turn}} \in \{2,3,4,5,21\}$ share the Model A materials and circuit topology, only the geometry varies depending on the turn count, increasing conductor length. To isolate the conditioning driven by the scalar potential block, the magneto-static problem with current excitation is solved as well as the eddy current problems.

#### 3) Model B: 21-turn coil model with skin-effect mesh

This model extends Model A by adding a finer mesh in the conductor regions to capture high-frequency skin effects as shown in Fig. 3. It is excited by a 1 V voltage source at 30 kHz. The mesh size increases with 3,670,328 edge DOFs in $\Omega$ and 514,650 node DOFs in $\Omega_c$.

#### 4) Model C: 18-turn independent conductor model

This model represents a different case involving 18 independently defined conductors forming a one-fourth symmetric coil geometry as shown in Fig. 4. Each conductor is excited independently with a 1 A current source at 30 kHz. Due to the symmetry assumption, 18 input and 18 output ports are

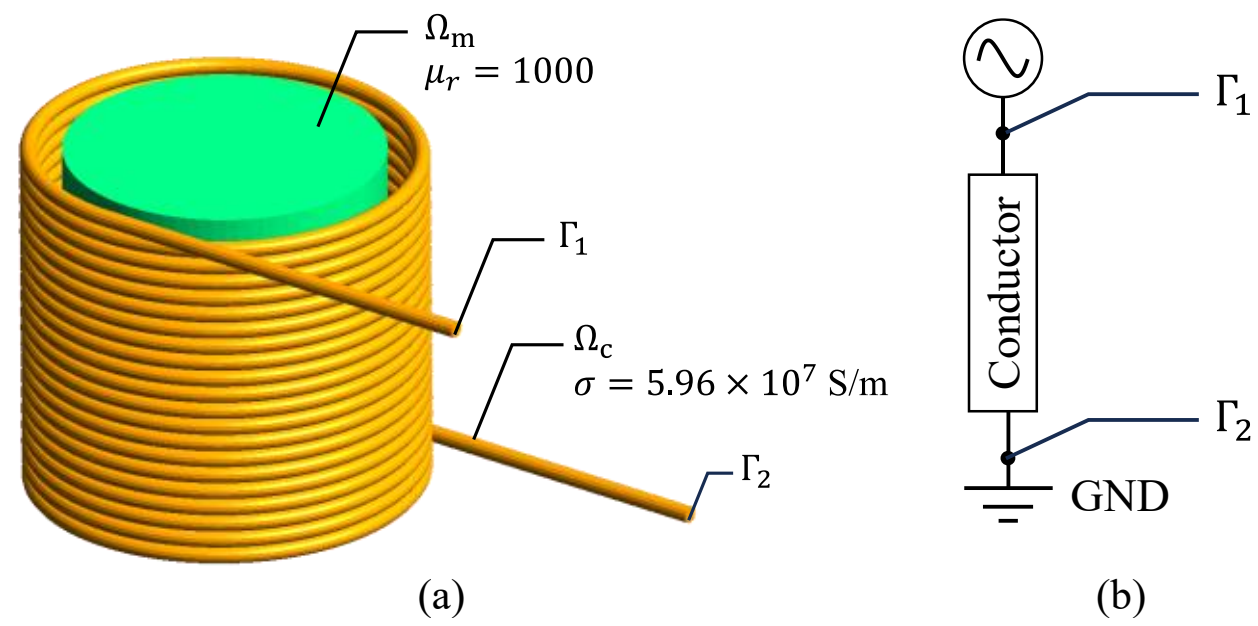

Fig. 2 (a) Geometry of the 21-turn coil model (Models A and B). (b) Equivalent external circuit with voltage excitation.

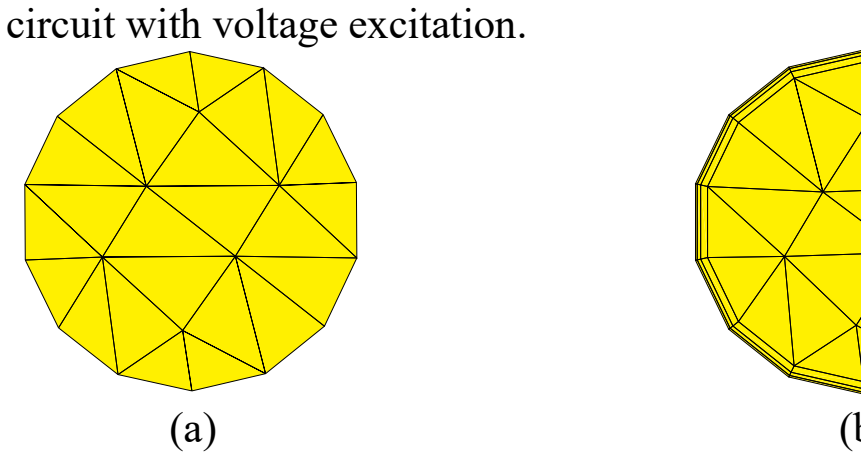

Fig. 3 Cross-sectional views of coil mesh configurations. (a) Coarse mesh used in Model A. (b) Fine skin-effect-resolving mesh used in Models B and C.

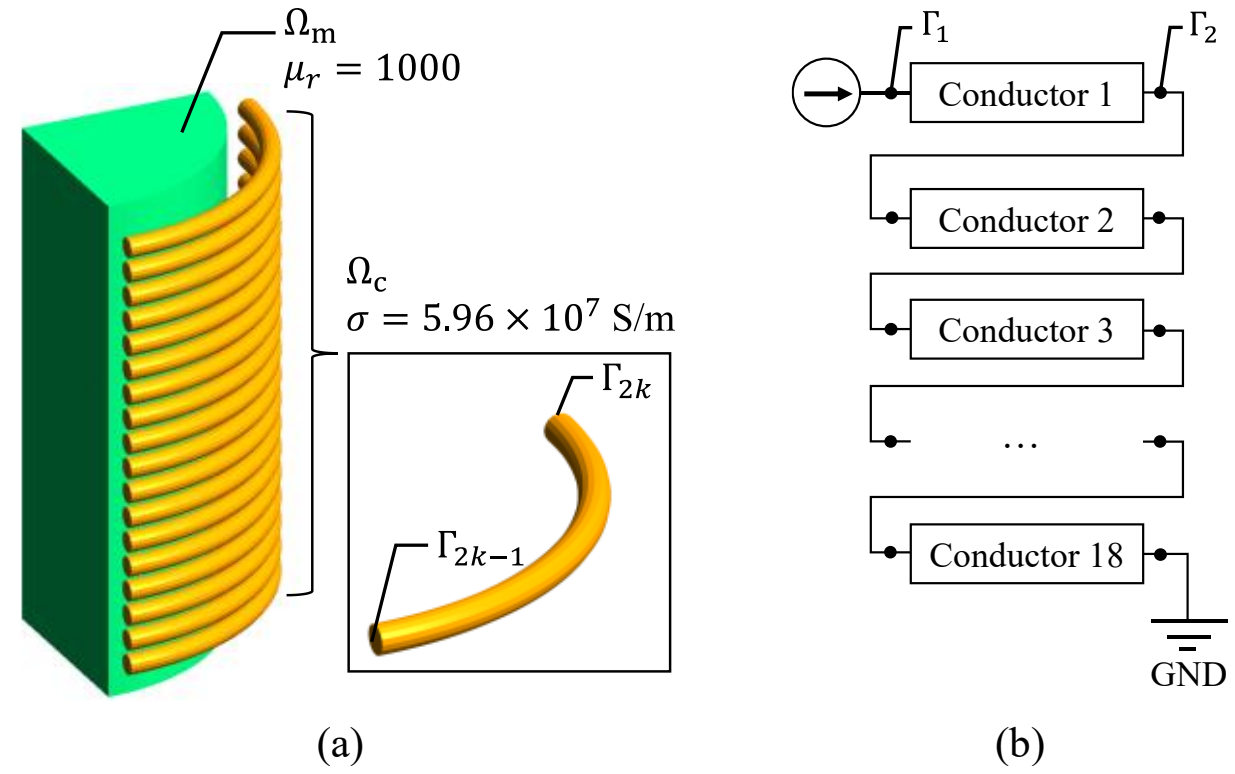

Fig. 4 (a) Geometry of the 18-turn independent conductor model (Model C). (b) Equivalent external circuit with current excitation.

defined. The model includes 828,300 edge DOFs in $\Omega$ and 129,445 node DOFs in $\Omega_c$.

Each model was constructed to investigate specific features of the preconditioner: Model A for baseline performance at low-frequency, Model $A_p$ for evaluating robustness against geometry-dependent spectral degradation, Model B for evaluating robustness against high-frequency configuration and large-scale model, and Model C for assessing the effectiveness of physics-level parallelization across independent conductor regions.

### B. Experimental Setup

All simulations were performed on a workstation equipped with Intel(R) Xeon(R) Gold 6244 CPU (3.60 GHz, 8 cores) and 192 GB of DDR4-2933 memory (6 × 32 GB modules). The parallelization was implemented using OpenMP shared-memory threading. Each computation used up to 8 threads.

The following preconditioning configurations were tested:

- IC: Incomplete Cholesky preconditioner with a diagonal shift parameter $\alpha = 1.1$ (baseline) [10]
- EMD (IC + AMG V-cycle): EMD preconditioner with IC for $M_1$ and V-cycle AMG for $M_2$
- EMD (IC + AMG W-cycle): EMD preconditioner with W-

TABLE I
PERFORMANCE METRICS FOR MODEL A (21-TURN COIL AT 50 HZ)

| Preconditioner | #Subdomains | #Iterations | Solver time [s] | Per-iteration time [s] | Speed-up ratio |
|---|---|---|---|---|---|
| IC (baseline) | - | 5949 | 452.5 | 0.0760 | 1.00 |
| EMD (IC + AMG V-cycle) | - | 288 | 21.8 | 0.0755 | 20.8 |
| EMD (IC + AMG W-cycle) | - | 267 | 24.2 | 0.0906 | 18.7 |
| EMD (IC + GenEO-DDM) | 8 | 356 | 45.0 | 0.1264 | 10.0 |

TABLE II
PERFORMANCE METRICS FOR MODEL $A_p$ (PARAMETRIC ELONGATED COIL AT 0 HZ AND 50 HZ).

| #Turns | #Iterations | | | Ratio: (eddy current, IC) / (magneto-static, IC) | Ratio: (eddy current, EMD) / (magneto-static, IC) |
|---|---|---|---|---|---|
| | Magneto-static | Eddy current with external circuit coupling | | | |
| | IC | IC | EMD (IC + LU) | | |
| 2 | 127 | 822 | 114 | 6.47 | 0.898 |
| 3 | 139 | 1094 | 124 | 7.87 | 0.892 |
| 4 | 147 | 1454 | 133 | 9.89 | 0.905 |
| 5 | 172 | 1711 | 151 | 9.95 | 0.878 |
| 21 | 329 | 5949 | 273 | 18.08 | 0.830 |

TABLE III
PERFORMANCE METRICS FOR MODEL B (21-TURN COIL WITH SKIN-EFFECT-RESOLVING MESH AT 30 KHZ).

| Preconditioner | #Subdomains | #Iterations | Solver time [s] | Per-iteration time [s] | Speed-up ratio |
|---|---|---|---|---|---|
| IC (baseline) | - | 15838 | 5964.8 | 0.377 | 1.00 |
| EMD (IC + AMG V-cycle) | - | 4069 | 1716.8 | 0.422 | 3.47 |
| EMD (IC + AMG W-cycle) | - | 2935 | 1552.9 | 0.529 | 3.84 |
| EMD (IC + GenEO-DDM) | 8 | 1106 | 654.2 | 0.591 | 9.12 |
| EMD (IC + GenEO-DDM) | 16 | 1138 | 593.1 | 0.521 | 10.06 |
| EMD (IC + GenEO-DDM) | 24 | 1004 | 550.8 | 0.549 | 10.83 |

TABLE IV
PERFORMANCE METRICS FOR MODEL C (18-TURN INDEPENDENT CONDUCTORS AT 30 KHZ).

| Preconditioner | Parallelization | #Iterations | Solver time [s] | Per-iteration time [s] | Speed-up ratio |
|---|---|---|---|---|---|
| IC (baseline) | Algorithm-level | 2476 | 125.9 | 0.0509 | 1.00 |
| EMD (IC + LU) | Physics-level | 776 | 72.4 | 0.0933 | 1.74 |
| EMD (IC + LU) | Algorithm-level | 776 | 385.0 | 0.4961 | 0.33 |

cycle AMG for $M_2$

- EMD (IC + GenEO-DDM): EMD preconditioner with DDM using generalized eigenproblems in the overlaps (GenEO) coarse space method for $M_2$ [5]
- EMD (IC + LU): EMD preconditioner with a direct solver applied to $M_2$

All solvers used the conjugate orthogonal conjugate gradient (COCG) method and all test cases were performed using double-precision arithmetic. The numerical implementation utilizes the PARDISO [11] library as a sparse direct solver for GenEO-DDM, and the AMGCL [12] framework as AMG and IC preconditioners. These components were selected for their robustness and compatibility with large-scale electromagnetic simulations and were incorporated into the EMD framework accordingly.

Finally, the following convergence criteria were considered:

- In voltage excitation problems (Models A, $A_p$ and B), the stopping condition was set to a relative residual norm of $\|r\|_2/\|b\|_2 < 10^{-10}$ defined from (8).
- In current excitation problems (Model C), the stopping condition was set to a relative residual norm of $\|r\|_2/\|b\|_2 < 10^{-8}$ defined from (7) and (8).

### C. *Results and Discussion*

The performance of various preconditioners was evaluated across the three benchmark models described in Section III-A. Tables I–IV summarize the iteration count, iterative solver time, per-iteration time and speed-up ratio for each configuration. Convergence histories of Model A are illustrated in Fig. 5.

The numerical results demonstrate the clear effectiveness of the proposed EMD preconditioner across various eddy current models. In Model A, a low-frequency configuration, EMD with IC + AMG V-cycle achieved the shortest iterative solver time among all tested methods. Notably, it also exhibited the lowest per-iteration cost, outperforming even the conventional IC preconditioner. This advantage can be attributed to the block-diagonal structure inherent in the EMD preconditioner. While the IC preconditioner operates on the entire system matrix, the EMD method decouples the vector and scalar components, allowing each to be preconditioned independently. The resulting preconditioner matrix is sparser. Although AMG generally has higher computational cost per iteration and lower parallel efficiency compared to IC, the reduced number of nonzeros due to the block decomposition compensates for this, leading to an overall reduction in both iterative solver and per-iteration time.

In Model $A_p$, increasing the number of turns raises the effective conductor length while keeping local cross-sectional scales unchanged. Under this change, iterations of the magneto-static with IC grow moderately. In contrast, for the eddy current problems with IC, the iteration count increases sharply with the number of turns, and the iteration ratio relative to the magneto-static baseline rises from 6.47 to 18.08. This monotone growth is consistent with the deterioration of the spectrum associated with the discrete Laplacian operator $G^\top M_\sigma G$. Applying EMD (IC + LU) restores robustness across all cases: the eddy-current iteration counts remain close to the magneto-static baseline. These results confirm that EMD effectively neutralizes the elongation-induced ill-conditioning while preserving favorable convergence behavior as the conductor length increases.

Model B highlighted the robustness of the method under high-frequency conditions with fine skin-effect-resolving meshes. While the conventional IC preconditioner required over 15,000 iterations, EMD with AMG and DDM preconditioners reduced the iteration count by more than an order of magnitude. The iterative solver time was up to 10 times faster. At low frequency, AMG-based preconditioners exhibit superior efficiency, whereas DDM-based approaches become more advantageous in higher-frequency regimes.

In Model C, which features multiple independently excited conductors, the effect of physics-level parallelization became evident. By exploiting the block-diagonal structure of the continuity equation across conductor regions, the preconditioner $M_2$ was independently applied to each region. This yielded nearly 5 times speed-up in iterative solver time when compared to an algorithm-level parallelization of $M_2$ where the forward elimination and backward substitution are parallelized in the algorithm level. Note that the per-iteration time increases because each application of $M_2$ invokes triangular solves from the LU factors; the total speed-up in physics-level parallelization is therefore modest. However, the important point is that physics-level parallelization allows each block to maintain complete independence while simultaneously using algorithm-level parallelization, which is important from the perspective of utilizing computational resources in large-scale calculations.

Overall, the experiments confirm that the EMD framework enables substantial improvements in both convergence and scalability. It provides a practical alternative to conventional IC-based solvers, especially for problems involving coupling with external circuits or complex conductor topologies.

## IV. Conclusion

This study proposed a novel block-structured preconditioning framework, termed EMD, for solving eddy current problems involving coupled external circuits. The key idea is to separately precondition the vector and scalar potential components in the discrete system, thereby addressing the spectral degradation induced by the discrete Laplacian.

Although the effectiveness of the proposed method was demonstrated under the A$-\varphi$ formulation for time-harmonic regime, several directions remain for future work. First, the applicability of the EMD concept should be examined for transient regime including nonlinearity and hysteresis in silicon steel sheet, as well as for alternative formulations such as the T$-\Omega$ method. Moreover, the current study focused on local voltage and current excitations; extending the framework to handle non-local excitation types, such as distributed sources or coupled systems, requires further investigation. In principle, the core idea of EMD remains promising for these cases because the underlying algebraic ill-conditioning persists. Finally, in practical scenarios involving a large number of conductors connected in series and parallel, the development of efficient preconditioning strategies that account for such electrical connectivity is another important research direction.

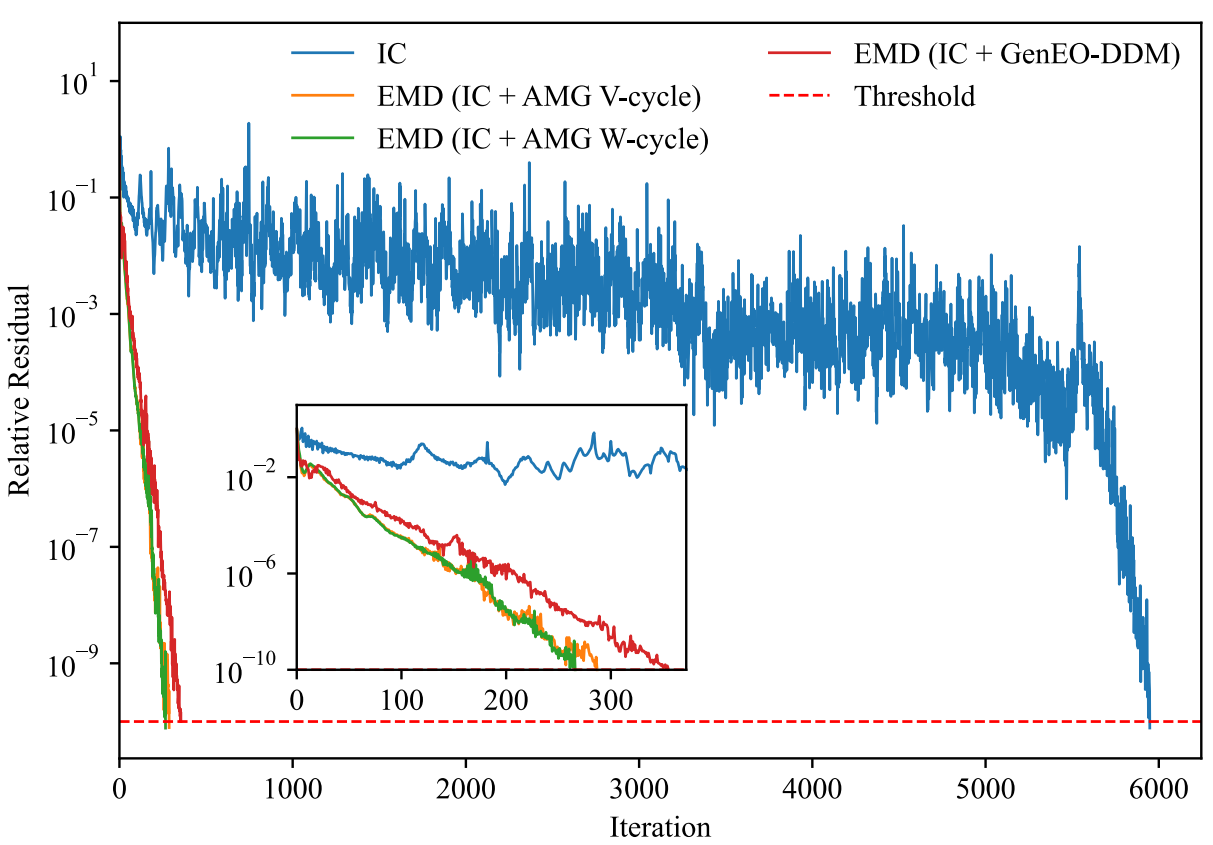

Fig. 5 Convergence histories for Model A (low-frequency 21-turn coil). Comparison of IC and EMD preconditioners in terms of residual norm versus iteration.

## Acknowledgment

This work was supported in part by JSPS KAKENHI Grant Number JP24K17261.

## References

[1] R. Barrett, M. Berry, T. F. Chan, J. Demmel, J. Donato, J. Dongarra *et al.*, *Templates for the solution of linear systems: building blocks for iterative methods*, SIAM, 1994.

[2] J. W. Ruge and K. Stüben, “Algebraic multigrid,” in *Multigrid Methods*, SIAM, 1987, vol. 3, pp. 73–130.

[3] S. Reitzinger and J. Schöberl, “An algebraic multigrid method for finite element discretizations with edge elements,” *Numer. Linear Algebra Appl.*, vol. 9, pp. 223–238, 2002.

[4] R. Hiptmair and J. Xu, “Auxiliary space preconditioning for edge elements,” *IEEE Trans. Magn.*, vol. 44, no. 6, pp. 938–941, Jun. 2008.

[5] N. Spillane, V. Dolean, P. Hauret, F. Nataf, C. Pechstein, and R. Scheichl, “Abstract robust coarse spaces for systems of PDEs via generalized eigenproblems in the overlaps,” *Numer. Math.*, vol. 126, no. 4, pp. 741–770, 2014.

[6] N. Bootland, V. Dolean, F. Nataf, and P. H. Tournier, “A robust and adaptive GenEO-type domain decomposition preconditioner for H(curl) problems in three-dimensional general topologies,” *arXiv preprint*, arXiv:2311.18783, 2023.

[7] R. Hiptmair and O. Sterz, “Current and voltage excitations for the eddy current model,” *Int. J. Numer. Model. Electron. Netw. Devices Fields*, vol. 18, no. 1, pp. 1–21, 2005.

[8] M. A. Al-Gwaiz, *Sturm-Liouville Theory and Its Applications*. London, U.K.: Springer London, 2008.

[9] R. Hiptmair, “Multigrid method for Maxwell’s equations,” *SIAM J. Numer. Anal.*, vol. 36, pp. 204–225, Dec. 1998.

[10] K. Fujiwara, T. Nakata, and H. Fusayasu, “Acceleration of convergence characteristic of the ICCG method,” *IEEE Trans. Magn.*, vol. 29, no. 2, pp. 1958–1961, Mar. 1993.

[11] O. Schenk, K. Gärtner, W. Fichtner, and A. Stricker, “PARDISO: A high-performance serial and parallel sparse linear solver in semiconductor device simulation,” *Future Gener. Comput. Syst.,* vol. 18, no. 1, pp. 69–78, 2001.

[12] D. Demidov, “AMGCL: An efficient, flexible, and extensible algebraic multigrid implementation,” *Lobachevskii Journal of Mathematics*, vol. 40 no. 5, pp535–546, May 2019.